\documentclass[12pt,setstack]{iopart}
\usepackage{epsfig}
\usepackage{epsf}
\usepackage{float}
\usepackage{cite}
\usepackage{graphicx}
\usepackage{iopams}

\def\ep{{\epsilon}}
\def\cg{{\cal G}}

\def\om{{\omega}}

\def\nnu{{\nonumber}}

\def\g{{\bf{g}}}

\def\P{{\bf{P}}}

\def\beq{\begin{equation}}
\def\eeq{\end{equation}}
\def\beqa{\begin{eqnarray}}
\def\eeqa{\end{eqnarray}}

\def\g0{{\gamma_0}}

\def\Im{{\mbox{Im}}}

\def\bear{\begin{eqnarray}}
\def\eear{\end{eqnarray}}
\def\bea{\begin{align}}
\def\eea{\end{align}}

\eqnobysec
\begin{document}
\bibliographystyle{plain}
\input epsf

\title[Magnetoresistance in paramagnetic heavy fermion metals]{Magnetoresistance in paramagnetic heavy fermion metals.}
\author{D.\ Parihari\dag, N.\ S.\ Vidhyadhiraja\ddag}
\address{\dag\ Department of Physics, Indian Institute of Technology
Kharagpur, \\ Kharagpur 721302, India.}
\address{\ddag\ Theoretical Sciences Unit,Jawaharlal Nehru Centre For
Advanced Scientific Research, Jakkur, Bangalore 560064, India}

\date{\today}

\begin{abstract}
 A theoretical study of magnetic field (h) effects on single-particle
spectra and transport quantities of heavy fermion metals in the paramagnetic
phase is carried out. We have employed a non-perturbative local moment 
approach (LMA)
to the asymmetric periodic Anderson model within the dynamical mean field
framework. The lattice coherence scale $\om_L$, which is proportional within
the LMA to the spin-flip energy scale, and has been shown in earlier studies
to be the energy
scale at which crossover to single impurity physics occurs, 
increases monotonically with increasing magnetic field. 
The many body Kondo resonance in the density of states
at the Fermi level splits into two with the splitting being proportional to
the field itself. For h$\geq$ 0, we demonstrate adiabatic continuity
from the strongly interacting case to a corresponding non-interacting limit, 
thus establishing Fermi liquid behaviour for heavy fermion
metals in the presence of magnetic field. In the Kondo lattice regime,
the theoretically computed magnetoresistance is found to be negative
in the entire temperature range.  We argue that such a result could be 
understood at $T\gtrsim \om_L$ by field-induced suppression of spin-flip 
scattering and at $T\lesssim \om_L$ through lattice coherence.
The coherence peak
in the heavy fermion resistivity diminishes and moves to higher temperatures
with increasing field. Direct comparison
of the theoretical results to the field dependent resistivity measurements 
in CeB$_6$ yields good agreement.
\end{abstract}
\pacs{71.27.+a,71.28.+d,71.30.+h,75.20.Hr}

\section{Introduction}
The investigation of lanthanide/actinide based heavy fermion(HF) systems
has been a central theme in condensed matter physics both
theoretically and experimentally~\cite{grewe}.
Their behaviour is quite distinct from conventional clean metals,
the basic physics being driven by strong spin-flip
scattering from essentially localized f-levels, generating the large
effective mass. 
The periodic
Anderson model(PAM) forms the general paradigm within which these
materials are studied. The minimal model consists
of a regular array of sites, each associated with a localized, non-degenerate
f-electron core orbital, coupled to a
delocalized conduction electron orbital via a local hybridization.
Neighboring conduction electron orbitals are connected via a 
hopping matrix element and electron interactions enter the
model via an on-site coulomb repulsion.

The dynamical mean field theory
(DMFT)~\cite{vollhard} has proved to be a very
powerful framework for studies of various lattice models such as the
Hubbard model or the PAM.
Within DMFT, which is exact in the limit of infinite dimensions, the 
self-energy
becomes spatially local or momentum independent. As a consequence,
lattice models
map onto an effective single-impurity Anderson model (SIAM) with a self 
consistently determined host~\cite{vollhard}.

The PAM in the absence of a magnetic field has been studied extensively within
the framework of DMFT. 
Some of the numerical or semi-analytical methods (impurity solvers) 
that exist for solving the
the effective SIAM that arises within DMFT are quantum Monte Carlo~\cite{qmc},
numerical renormalization group~\cite{nrg}, iterated perturbation 
theory~\cite{ipt},
noncrossing approximation~\cite{nca}, the local moment 
approach~\cite{lma_hubb,raja_epjb}, 
the large-N/slave-boson~\cite{slave}, exact diagonalization~\cite{ed}
self-consistent perturbation theory~\cite{pert}, the Gutzwiller variational
method~\cite{gva} and the average T-matrix apprximation~\cite{tma}.
Every method has its own advantage, however
most of them suffer from one or the other limitations. For example,
the QMC suffers from minus sign problem at low temperature or large 
interactions, while IPT is able to capture only an algebraic decay of the Kondo
scale with increasing interactions. ED and NRG are in principle exact,
as is QMC, but since the spectral functions are obtained as a set of discrete
poles, a broadening is required, which is non-uniquely specified.

In this context, the
(LMA)\cite{logan,logan1,logan3,logan4a,logan4b,raja_epjb},
a diagrammatic theory based non-perturbative
many body method, has emerged as an approach for the single
impurity model and even for the lattice models within DMFT, that overcomes
some of the limitations mentioned above.
In particular, one obtains the exact dependence on interactions and 
hybridization of the Kondo scale of the
SIAM in strong coupling as the Bethe ansatz solution for the Kondo 
model~\cite{logan}.  Very good agreement has been seen between LMA and NRG 
results for the spectral functions of the SIAM~\cite{logan4a,logan4b}. Within DMFT, the
LMA has proved quite successful in describing 
the spectral and transport properties
of several metallic and insulating heavy fermion systems such as CeB$_6$,
SmB$_6$, YbAl$_3$, CeAl$_3$, YbB$_{12}$, CeOs$_4$Sb$_{12}$ in the 
paramagnetic phase~\cite{logan9,rajahvtheo,rajahvexpt}. 
The LMA is computationally inexpensive, yields quantities in
real frequency directly, and is semi-analytical. One limitation
of LMA is that it is based on a symmetry restoration ansatz, that is
not easily generalizable to other problems, such as the multi-orbital
or multi-channel cases. Recently, a generalization of LMA has been reported
for the multi-orbital Anderson and Hubbard model~\cite{vlma}. Further,
it was shown that the LMA is a conserving approximation.
The other limitation is that since LMA is an approximate theory, one has
to benchmark its results against more exact theories such as NRG and QMC
to ascertain its reliability. Nevertheless, given the several advantages
above, and benchmarks, the LMA within DMFT is an appropriate choice to
study the effects of magnetic fields in heavy fermions.

Several theoretical studies of the effects of magnetic field on 
heavy fermion systems using either the Kondo lattice model (KLM) or the PAM
have been reported. The magnetic field induced insulator-metal transition
in Kondo insulators has been studied by various groups~\cite{saso,beach0,dp1}.
As relevant to heavy
fermion metals, the metamagnetic transition has been studied
using large-N mean-field~\cite{beach1} and subsequently
using DMFT+QMC~\cite{beach2}. 
In this work, our objective is to understand the magnetotransport
in heavy fermion metals. Previous work in this direction has been carried
out mainly either using the single-site Anderson models~\cite{graf}, 
thus missing out the lattice coherence effects completely or using large-N
mean field treatment of the Anderson lattice Hamiltonian~\cite{chen}. In 
order to
capture the lattice coherence effects
along with the single-impurity incoherent regime quantitatively within
a single framework, we employ the finite-field
LMA~\cite{logan4a,logan4b,dp1} within DMFT
for the periodic Anderson model away from half-filling and determine the
effect of magnetic field on spectra and transport of heavy fermion metals.
Our focus has been
on the strong coupling regime, and the quantities that we have studied
are spectral functions and magnetoresistance.
 The paper is organized as follows; the model and formalism are presented
in section II, followed, in section III by results for the field evolution
of single-particle dynamics and
 d.c transport. In section IV, experimental magnetoresistance of
 $CeB_{6}$ is compared with theory; and finally we
 conclude with a brief summary.

\section{Model and formalism}

The Hamiltonian for the PAM in standard notation is given by: 
\beq
\fl
\hat{H}=-t\sum_{(i,j),\sigma}c^{\dagger}_{i\sigma}
c^{\phantom{\dagger}}_{j\sigma} 
+ \sum_{i\sigma}(\epsilon_{f}+\frac{U}{2}f^{\dagger}_{i-\sigma}
f^{\phantom{\dagger}}_{i-\sigma})
f^{\dagger}_{i\sigma}f^{\phantom{\dagger}}_{i\sigma} 
+ V \sum_{i\sigma}
(f^{\dagger}_{i\sigma}c^{\phantom{\dagger}}_{i\sigma}+h.c) 
+
\sum_{i\sigma}\epsilon_{c}c^{\dagger}_{i\sigma}c^{\phantom{\dagger}}_{i\sigma}
\label{eq:pam}
\eeq
The first term describes the kinetic energy of the noninteracting
conduction ($c$) band due to nearest neighbour hopping $t$.
The second term refers to the $f$-levels with
site energies $\epsilon_{f}$ and on-site repulsion
$U$, while the third term describes the $c/f$ hybridization via the
local matrix element $V$. The final term represents the $c$-electron
orbital energy. In the limit of large dimensions, the hopping
needs to be scaled as $t\propto t_*/\sqrt{Z}$, where $Z$ is the
lattice coordination number. 
We consider the hypercubic lattice, for-which 
the non-interacting density of states is an unbounded Gaussian 
($\rho_{0}(\epsilon)= exp(-\epsilon^{2}/t^{2}_{*})/\sqrt(\pi t_{*})$).
 Particle-hole asymmetry in the PAM could be introduced in two ways
\cite{raja_epjb}: (i) through an asymmetric conduction 
band, {\it i.e.} $\epsilon_{c} \neq 0$, or (ii) through an asymmetric f-level,
$\epsilon_{f}\neq -\frac{U}{2}$. In general, we may quantify the 
f-level asymmetry by defining $ \eta=1+\frac{2\ep_{f}}{U}$,
such that $\eta=0$ is equivalent to particle-hole symmetric f-levels. 
Our primary interest is in the strong coupling Kondo lattice regime ($n_{f}
\rightarrow 1$) but with arbitrary conduction band filling ($n_{c}$).

Within DMFT, the PAM may be
mapped onto an effective self-consistent impurity problem within
DMFT. We choose the local moment approach to solve the 
effective impurity problem arising within DMFT. For details
of LMA developed for use within DMFT in the absence of a field, the 
reader is referred to
some of our previous work~\cite{logan9,raja_epjb,rajahvtheo,rajahvexpt}.
Magnetic field effects in the symmetric PAM as appropriate 
to Kondo insulators
was recently studied by us ~\cite{dp1}.  As discussed in that work, 
the presence of a global magnetic field results in the Zeeman splitting of
the bare electronic energy levels as $\epsilon_{\gamma\sigma}
=\epsilon_{\gamma}-\sigma h_\gamma$,
for $\gamma = c$ and $f$ electrons. 
Here $h_\gamma=\frac{1}{2}g_{\gamma}\mu_{B}H$ and
$\mu_{B}$ is the Bohr magneton; the constants $g_{f}$ and $g_{c}$
are the electronic g-factors for the $f $ and $c$ electrons respectively.
Although $g_{f}\neq g_{c}$ in general, for simplicity we set 
$g_{f} = g_{c}$.
The degeneracy of the symmetry broken solutions at the mean-field level 
(denoted by A and B) is lifted by a magnetic field. 
We consider here $h > 0$ for which $+|\mu(h)|$ (A-type) is the
sole solution.  The Feenberg self energy~\cite{logan9,raja_epjb,dp1}
becomes a functional solely
of the A-type Green's function,
{\it i.e.\ }$S_{\sigma}(\om)\equiv S_\sigma[G_{A\sigma}^{c}]$. Since the
B-type solution does not exist, the label `A' will be implicit in the 
following.

In presence of a uniform magnetic field, the LMA Green functions 
are given by
\beqa
\fl
G_{\sigma}^{c}(\omega,T,h)=\left[\omega^{+}+\sigma h-\epsilon_{c}
-S_{\sigma}^{c}[G_{\sigma}^{c}] 
-\frac{V^{2}}
{\omega^{+}+\sigma h-\epsilon_{f}-\tilde\Sigma_{\sigma}(\omega,T,h)}
\right]^{-1}
\label{eq:gcasy}\\
\fl
G_{\sigma}^{f}(\omega,T,h)=\left[\omega^{+}+\sigma h-\epsilon_{f}-
\tilde\Sigma_{\sigma}(\omega,T,h) 
- \frac{V^{2}}
{\omega^{+}+\sigma h-\epsilon_{c}-S_{\sigma}^{c}[G_{\sigma}^{c}]}\right]^{-1}
\label{eq:gfasy}
\eeqa
with $G^{\gamma}(\omega)=\frac{1}{2}\sum_{\sigma}G_{\sigma}^{\gamma}
(\omega)$. The self energy can be written as a sum
of the static and dynamic parts as,
\beq
\tilde\Sigma_{\sigma}(\omega,T,h)=\frac{U}{2}(\bar{n}-
\sigma|\bar{\mu}|(T,h))+
\Sigma_{\sigma}(\omega,T,h)
\label{eq:totalselfasy}
\eeq
where $|\bar{\mu}|(T,h)$ is the UHF local moment in presence of 
magnetic field and temperature. The dynamical self-energy within the LMA
is given by~\cite{logan3,dp1}
\beqa
\fl
\Sigma_{-\sigma}(\omega,T,h)& = U^{2}\int_{-\infty}^{\infty} 
\int^{\infty}_{-\infty}\,d\omega_1\,d\omega_{2}\,
\chi^{-\sigma\sigma}(\omega_{1},T,h)\;
 \frac{{\cal{D}}_{\sigma}(\omega_{2},h)}
{\omega^+ +\omega_{1}-\omega_{2}}\; h(\omega_1;\omega_2) 
\label{eq:selfasy}\\
\fl h(\omega_1;\omega_2)&=\theta(-\omega_1)[1-f(\omega_2,T)]+\theta(\omega_1)
f(\omega_2,T)\nnu
\eeqa
where $\chi^{-\sigma\sigma}(\omega,T,h) = \pi^{-1}{\rm Im}\Pi^{-\sigma\sigma}
(\omega,h)$, 
$f(\omega)=[\exp^{\beta\omega}+1]^{-1}$ is the Fermi function and
$\theta(\om)$ is the Heaviside step-function. 
Here, $\Pi^{-\sigma \sigma}(\omega,T,h)$ denotes the 
transverse spin polarization propagator which can be expressed as
$\Pi^{+-} =\, ^{0}\!\Pi^{+-}/(1 - U\, ^{0}\!\Pi^{+-})$ where
$^{0}\!\Pi^{+-}$, the bare p-h bubble, is constructed using 
the field dependent mean-field spectral densities~\cite{logan,rajahvtheo}.
The host spectral function is given by ${\cal{D}}_\sigma(\omega,h)=
-\frac{1}{\pi}{\rm Im}\cg_\sigma(\omega,h)$;
where the host/medium Green's function  $\cg_\sigma$ is given 
by,
\beq
\cg_\sigma(\omega,T,h) = \left[ \omega^{+}-e_f+\sigma x +\sigma h 
-\frac{V^{2}}
{\omega^{+}-\epsilon_c+\sigma h-S^{c}(\omega,T,h)}\right]^{-1}
\label{eq:crgasy}
\eeq
where the parameters $x=U|\mu|/2$ and $e_f$ are determined at 
$h,T=0$ by satisfying the symmetry restoration condition 
\beq
\Sigma^{R}_{\uparrow}(\omega=0; e_{f},x)-\Sigma^{R}_{\downarrow}
(\omega=0; e_{f},x)=U|\bar{\mu}(e_f,x)|.
\label{eq:srasy}
\eeq
and the  Luttinger's integral theorem\cite{raja_epjb,luttinger}
\beq
I_{L}(e_f,x)=\Im\int^{0}_{-\infty}\frac{d\omega}{\pi}
\frac{\partial\Sigma(\omega)}{\partial\omega}G^{f}
(\omega)=0
\label{eq:lutt}
\eeq
which in turn ensures a Fermi liquid
ground state. Note that $e_f$ is distinct from $\ep_f$. The latter
is a `bare' model parameter while the former is a derived `shifted
chemical potential', which is determined through the imposition of the
Luttinger's integral, or equivalently the Friedel sum rule~\cite{raja_epjb}.
In practice however, it is more convenient numerically
to fix $x$ and $e_f$ at the outset and treat $U$ and $\ep_f$ as
unknown parameters to be determined by the above two 
conditions~\cite{raja_epjb}.
As input for the calculation at a given field/temperature, we use the 
self-energies
and Green's functions of a lower field/temperature.

In the limit of infinite dimensions, vertex corrections in the 
skeleton expansion for the current-current correlation function 
are absent, hence a knowledge of single-particle dynamics is sufficient 
within DMFT to determine $\bf{q}$=0 transport properties. For
$h>0$, the $d$-dimensional isotropic conductivity is computed by
adding the contributions from each of the spin channels as
\beq
\fl \bar{\sigma}(\omega,T,h)=\frac{\sigma_{0} t^{2}_{*}}{d\,\omega}\sum_{\sigma}
\int_{-\infty}^{\infty}d\omega_{1}[f(\omega_{1})- f(\omega_{1}+\omega)]
\; 
\langle D_{\sigma}^{c}(\epsilon;\omega_{1})D_{\sigma}^{c}(\epsilon;
\omega_{1}+\omega)\rangle_{\epsilon}
\eeq
where 
$\sigma_{0}=
\frac{\pi e^{2}}{\hbar a}$ (a is the lattice parameters) and
$D_{\sigma}^{c}(\epsilon;\omega)=-\frac{1}{\pi}{\rm Im}G_{\sigma}^{c}
(\epsilon;\omega)$. The lattice $c$-Green's function is given by
$G_{\sigma}^{c}(\epsilon;\omega)=[\gamma_{\sigma}(\omega,T,h)-\epsilon]^{-1}$
where 
\begin{displaymath}
\gamma_\sigma=\om^+ + \sigma h -\ep_c - V^2\left( \om^+ + \sigma h
-\ep_f -\tilde{\Sigma}_\sigma(\om,T,h)\right)^{-1}\,,
\end{displaymath}
 and the $\epsilon$ - average is defined by,
\beq
\langle A(\epsilon;\omega)\rangle_{\epsilon}=
\int^{\infty}_{-\infty}d\epsilon \rho_{0}(\epsilon)A
(\epsilon;\omega)
\eeq

We now proceed to discuss 
the results obtained by implementing the finite field and finite temperature
LMA.

\section{Results and discussions}

In this section, we will discuss the single particle dynamics and
transport in presence of magnetic field using LMA. Our primary
focus is on the strong coupling Kondo lattice regime 
(where $n_{f}\rightarrow$ 1, but $n_c$ is arbitrary).
Before discussing the finite field calculations, we will review a few well
established concepts. At $T=0$ and $h=0$\cite{raja_epjb},
the strong coupling Kondo lattice regime is
characterized by an exponentially small (in strong coupling) low- energy 
scale $\omega_{L}=ZV^{2}/t_*$. 
The single-particle properties of
the asymmetric PAM exhibit universal scaling in terms of $\omega/\omega_{L}$ 
and for
a fixed $\epsilon_{c}/t_*$ and $\eta$. 
For finite-$T$ and $h=0$,
the spectra
$D^{c}(\omega)$ and $V^{2}D^{f}(\omega)$ exhibit scaling
in terms of $\omega/\omega_{L}$ and $T/T_L (T_L=\omega_L$). 
In summary, the spectra and transport
properties of the PAM are universal functions of $\omega/\omega_{L}$ and
$T/\omega_L$, for a given conduction band filling and f-level asymmetry;
thus being independent of the bare $U/t_*$ and $V/t_*$. In this context, it is 
important to mention that the universal form of the scaling functions
does depend, albeit weakly, on the specific lattice, which manifests
in the bare conduction band density of states. 
The application of magnetic field does not destroy this universality
as shown in a recent
work on the symmetric PAM ($T=0$) where such scaling 
in terms of
$\tilde\omega_{h}=\omega t_*/(Z(h)V^{2})$ for a fixed effective
field (detailed description is in \cite{dp1}) was shown to
hold good in the presence of field as well. 

To see if such scaling occurs for the asymmetric case, we carry out
a low frequency Fermi liquid analysis of the Green's functions by
expanding the self-energy about
the Fermi level to first order in $\omega$ as,
\beq
\Sigma^{R}_{\sigma}(\omega,h)=\Sigma^{R}_{\sigma}(0,h)+
(1-\frac{1}{Z_{\sigma}(h)})\omega
\label{eq:self}
\eeq
Substituting equation~\eref{eq:self} in equations~\eref{eq:gcasy} and 
~\eref{eq:gfasy}, we find that the spin
dependent spectral functions are just renormalized versions of their
non-interacting counterparts and are given by(neglecting the
`bare' terms h and $\omega$ in the strong coupling limit),
\bear
D^c_\sigma(\omega;h)&\stackrel{\omega\rightarrow 0}{\rightarrow}&
\rho_0\left(-\epsilon_{c}-\frac{1}{\tilde\omega_{h\sigma} -
\tilde\epsilon^{*}_{f} + \sigma h^{\sigma}_{eff}}\right)
\label{eq:rnonc}
\eear
\bear
D^f_\sigma(\omega;h)&\stackrel{\omega\rightarrow 0}{\rightarrow}&
\frac{t^{2}_{*}}{V^{2}(\tilde\omega_{h\sigma} -\tilde\epsilon^{*}_{f} +
\sigma h^{\sigma}_{eff})^{2}}D^c_\sigma(\omega;h)
\label{eq:rnonf}
\eear
where $\tilde\omega_{h\sigma}=\omega t_{*}/(Z_{\sigma}(h)V^{2})$ and
the renormalized f-level ($\tilde\epsilon^{*}_{f}$) and $h^{\sigma}_{eff}$
can be expressed as,
\beq
\tilde\epsilon^{*}_{f}=\frac{t_{*}}{V^{2}}(\epsilon_{f}+\frac{U\bar{n}}{2}-
\frac{\sigma U\bar{\mu}}{2}+\Sigma^{R}_{\sigma}(0,0))
\label{eq:ref}
\eeq
and
\beq
h^{eff}_{\sigma}=\frac{t_{*}}{V^{2}}(h-\sigma(\Sigma^{R}_{\sigma}(0,h)-
\Sigma^{R}_{\sigma}(0,0)))
\label{eq:heff}
\eeq
Using the SR condition, equation~\eref{eq:srasy}, it is easy to see that
the effective $f$-level, $\tilde\epsilon^{*}_{f}$ is spin independent.
Note that the quantities thus defined, namely, $\tilde{\omega}_{h\sigma},
\tilde\epsilon^{*}_{f}$ and $h^{eff}_{\sigma}$ are dimensionless.

From equations~\eref{eq:rnonc} and ~\eref{eq:rnonf}, we can 
infer the following:
\noindent
(i)\ The spin dependent spectra $D^{c}_{\sigma}(\omega,h)$ and
$V^{2}D^{f}_{\sigma}(\omega,h)$ should exhibit scaling in terms
of $\tilde\omega_{h\sigma}$ for a fixed $h_{\sigma}^{eff}$ {\it i.e},
for a fixed $h_{\sigma}^{eff}$, if we plot the spin dependent spectra {\it vs}
$\tilde\omega_{h\sigma}$ for different values of 
U, they should collapse onto a single curve at the low energy regions
for a particular $h_{\sigma}^{eff}$.
\begin{figure}
\centerline{\hbox{
\epsfig{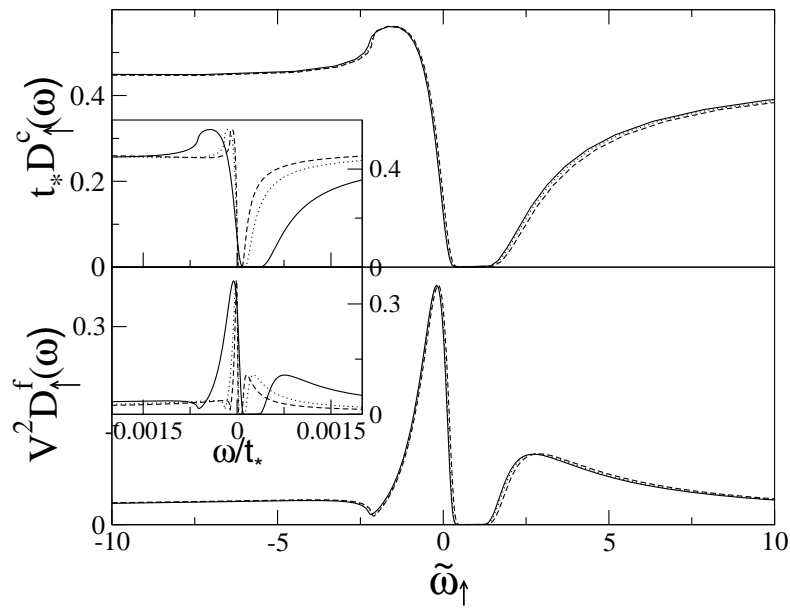}}}
\caption{Insets: LMA spin dependent conduction and f- electron spectra
{\it vs} bare frequency $\omega/t_{*}$ for $\epsilon_{c}=0.3t_*$,
$\eta \rightarrow$ 0, $V^{2}=0.2t_*^2$ and $T=0$ for
three parameter sets:
$U \sim  5.1t_*$ and $\tilde{h}=0.3$ (solid),
$U \sim  6.1t_*$ and $\tilde{h}=0.25$ (dotted), $U \sim  6.6t_*$,
and $\tilde{h}=0.24$ (dashed) for fixed $h_{\uparrow}^{eff}=0.39$
( where $\tilde{h}=ht_*/(Z(0)V^{2})$).
Main: The same spectra when plotted {\it vs} $\tilde\omega_{h\uparrow}=
\omega t_{*}/(Z_{\uparrow}(h)V^{2}$ collapse into one single universal form.}
\label{fig:univ1}
\end{figure}
This is shown in figure~\ref{fig:univ1} where in the insets,
the conduction electron spectra for up spin (top panel) and the 
f-electron spectra for up spin (bottom panel)
are plotted as a function of the bare frequency $\omega/t_{*}$
with  $\epsilon_{c}=0.3t_*$,
$\eta\rightarrow$ 0, $V^{2}=0.2t_*^2$ and T=$0$ for
three parameter sets:
$U \sim  5.1t_*$ and $\tilde{h}=0.3$ (solid),
$U \sim  6.1t_*$ and $\tilde{h}=0.25$ (dotted), $U \sim  6.6t_*$,
and $\tilde{h}=0.24$ (dashed) such that $h_{\uparrow}^{eff}$=0.39 is fixed
(where $\tilde{h}=ht_*/(Z(0)V^{2})$).
In the main panel the same
spectra are plotted as a function of  $\tilde{\omega}_{h\uparrow}$.
We see that when we plot the spectra {\it vs} bare frequency,
they appear very different, but when
plotted {\it vs.} $\tilde\omega_{h\uparrow}$ 
(main panel: top and
bottom), they  collapse onto a single universal form. We have checked that
similar universality holds for the down spin also for a fixed 
$h_{\downarrow}^{eff}$.

\begin{figure}[h]
\centerline{\hbox{
\epsfig{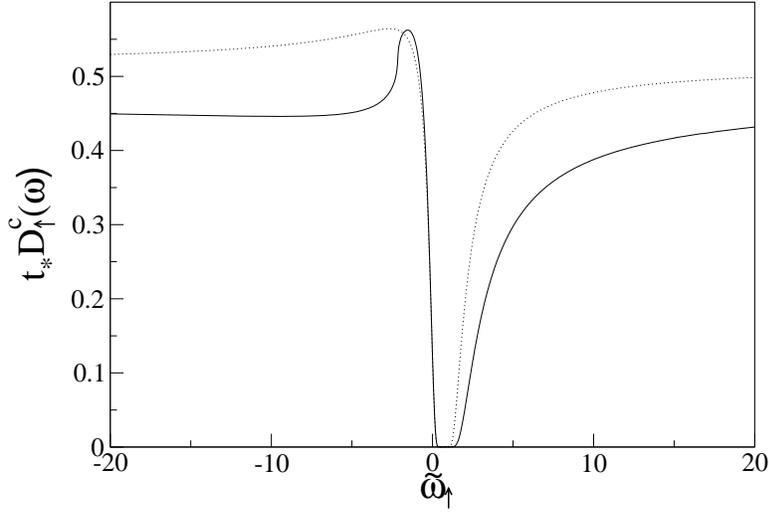} }}
\caption{LMA spin dependent scaling spectra (solid lines) are
superposed onto the
corresponding non-interacting spectra (dotted lines) for a fixed
$h_{\uparrow}^{eff}=0.45$ and $\tilde\epsilon^{*}_{f}=1.09$.}
\label{fig:adia}
\end{figure}

\noindent
(ii)\ The spin dependent conduction and f-electron spectra
$D^{c}_{\sigma}(\omega)$ and $V^{2}D^{f}_{\sigma}(\omega)$
should adiabatically connect
to the non-interacting limit at low energy scales {\it i.e}, if we
take the $c$-electron and $f$-electron fields as zero and
$h_{\sigma}^{eff}$ respectively,  substitute for $\ep_f$
the renormalized f-level ($\tilde{\epsilon}^{*}_{f}$)
and compute the spectra in the non-interacting limit; 
then the interacting 
spectra and the non-interacting spectra should be same at 
low energy scales. This is
shown in figure~\ref{fig:adia}, where the spin dependent scaling spectra
$D^{c}_{\uparrow}(\omega)$ for the interacting case ($U=6.6t_*$) is
superposed onto the non-interacting spectra. We 
see from figure~\ref{fig:adia} that both the curves are almost 
identical near the Fermi level. This demonstrates
adiabatic continuity of the strong coupling regime to the
non interacting limit which represents Fermi liquid behaviour
in presence of magnetic field for the asymmetric case (in parallel
to the symmetric PAM~\cite{dp1}).

\begin{figure}[t]
\centerline{\hbox{
\epsfig{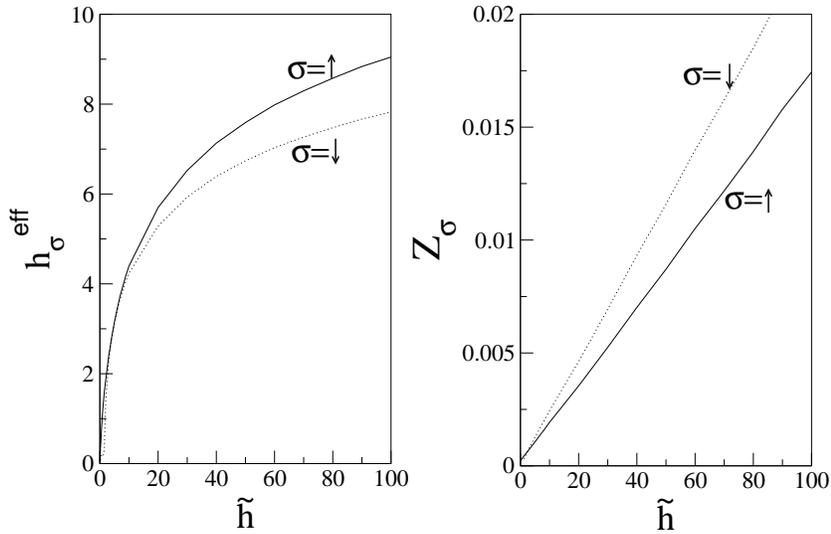}}}
\caption{Left panel: $h_{\sigma}^{eff}$ {\t vs}
$\tilde{h}=ht_*/(ZV^{2})$( $Z$ is the quasiparticle weight for $h=0$)
for $\epsilon_{c}=0.3t_*$, $\eta\rightarrow 0$, $T=0$ and $U = 6.6t_*$.
Right panel: Spin dependent quasiparticle weight $Z_{\sigma}(h)$ {\it vs}
$\tilde{h}$ for the same parameters.}
\label{fig:heff}
\end{figure}

\begin{figure}[t]
\centerline{\hbox{
\epsfig{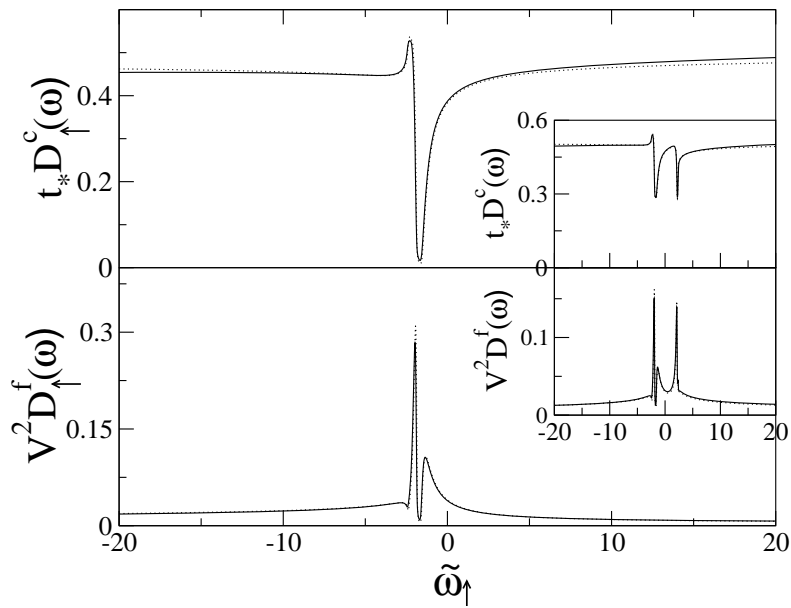}}}
\caption{Main panel: LMA up-spin conduction electron scaling
spectra (top) and up spin $f$-electron scaling spectra (bottom)
{\it vs} $\tilde\omega_{h\uparrow}$ for $\epsilon_{c}=0.3t_*$,
$\eta\rightarrow$ 0 and $V^{2}=0.2t_*^2$ for
two parameter sets:
$U \sim  5.1t_*$ and $\tilde{h}=10.0$ (solid),
$U \sim  6.6t_*$ and $\tilde{h}=9.0$ (dotted) for
fixed $h_{\uparrow}^{eff}=4.2$.
Insets: The corresponding spin summed conduction and (top) $f$-electron
(bottom) scaling spectra {\it vs} $\tilde\omega_{h\uparrow}$ for the
same parameter sets. Spin summed as well as the spin dependent spectra
exhibit universality/scaling in the strong coupling regime.}
\label{fig:univ2}
\end{figure}

To see the effect of magnetic field on the spin dependent
effective fields $h_{\sigma}^{eff}$ and the spin dependent quasiparticle 
weights $Z_{\sigma}$, we have plotted in figure~\ref{fig:heff}, 
the spin dependent effective fields
($h_{\uparrow}^{eff}$ and $h_{\downarrow}^{eff}$) {\it vs}
$\tilde{h}$ (left panel) and the spin dependent
quasiparticle weights ($Z_{\uparrow}(h)$ and $Z_{\uparrow}(h)$) {\it vs}
 $\tilde{h}$ (right panel). 
 We see that $Z_{\uparrow}(h)\cong 
 Z_{\uparrow}(h)$ and $h_{\uparrow}^{eff} \cong h_{\downarrow}^{eff}$ 
 up to about $\tilde{h}\sim 10.0$. So, we can assume that the 
spin summed spectra $D^{c}(\omega)$ and
$V^{2}D^{f}(\omega)$ also should exhibit scaling in
terms of $\tilde\omega_{h\sigma}$ for low fields 
(up to $\tilde{h} \sim 10$)  for a  fixed $h_{\sigma}^{eff}$. 
To see this, we plot up spin scaling spectra (main panel) as well as 
corresponding spin summed spectra in figure~\ref{fig:univ2} (insets) for 
$U=5.1t_*$ (solid line) and $U=6.6t_*$ (dotted line) for
a fixed $h_{\uparrow}^{eff}=4.2$ ($\tilde{h}\sim 10.0$) and $T=0$. 
Indeed, we see that the spin summed spectra also exhibit universal scaling 
and adiabatic continuity at low energy scales.
We wish to reiterate the dicussion in the beginning of this section that
that the universality and scaling demonstrated above holds for a fixed
conduction band filling $n_c$, a fixed $f$-level asymmetry and for a 
specific bare conduction band density of states. In other
words, the functional forms of the spectra shown above would change
for a different choice for $n_c$ or $\eta$ or $\rho_0(\ep)$.

\begin{figure}[t]
\centerline{\hbox{
\epsfig{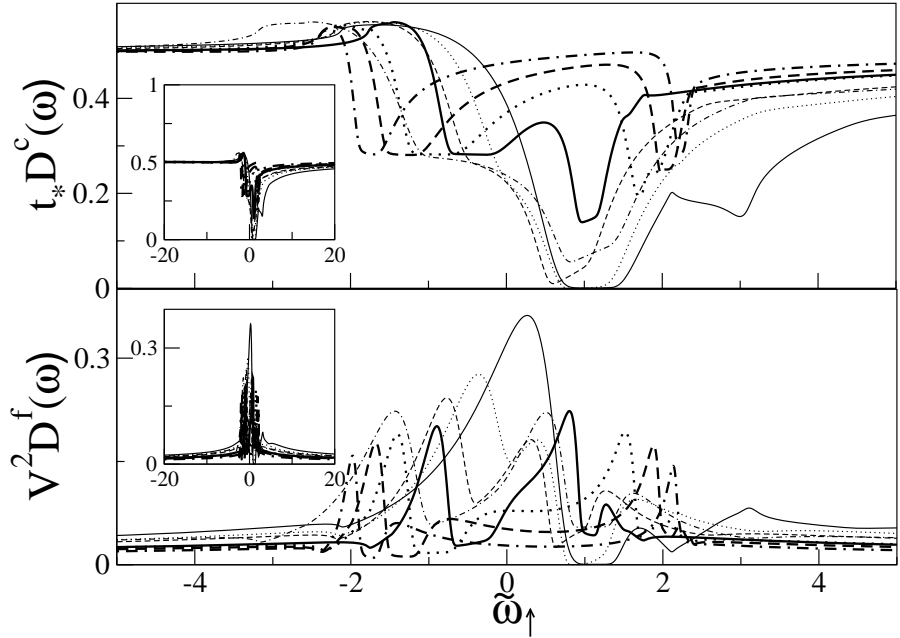}}}
\caption{LMA conduction and $f$- electron spectra {\it vs} 
$\tilde{\omega}_{h\uparrow}=
\omega\, t_*/(Z_{\uparrow}(h)V^{2})$ for $U=6.6t_*$, $\epsilon_{c}=0.3t_*$ and
$\eta\rightarrow 0.0$ for $\tilde{h}=0$ (solid), $\tilde{h}=
0.5$ (dotted),
1.0 (dashed), 1.5 (dot-dashed),
2.0 (bold solid), 3.0
(bold dotted), 5.0 (bold dashed) and
10.0 (bold dot-dashed) for $T=0$.}
\label{fig:graph1}
\end{figure}

\begin{figure}[h]
\centerline{\hbox{
\epsfig{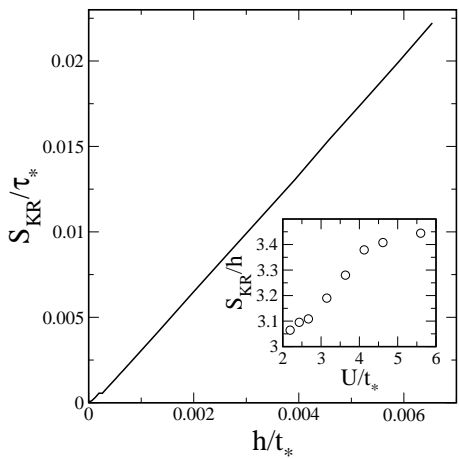}}}
\caption{Splitting of the Kondo resonance {\it vs} field for $U=5.5t_*$
and $\ep_c=0.5, V^2=0.2t_*^2$. The slope of the line in the main 
panel, {\it i.e.}
$S_{KR}/h$ as a function of interaction strength $U$ is shown in the inset.}
\label{fig:shiftdn}
\end{figure}

Now we turn to the LMA results for the field evolution of the asymmetric
PAM for a fixed temperature. In figure~\ref{fig:graph1}, we show $T=0$
 spin summed conduction and $f$-electron scaling spectra for
various fields: $\tilde{h}=0$ (solid), 
0.5 (dotted),
1.0 (dashed), 1.5 (dot-dashed),
2.0 (bold solid), 3.0
(bold dotted), 5.0 (bold dashed) and
10.0 (bold dot-dashed) and fixed interaction strength 
 $U=6.6t_*$. In our earlier work on the 
symmetric PAM~\cite{dp1}, we have shown that there is a gap at the Fermi level,
as there must be for a Kondo insulator.
But in the asymmetric limit $\epsilon_{c}=0.3t_*, h=0$
(see figure~\ref{fig:graph1}), the gap moves away from the Fermi level
and becomes a pseudo-gap. 
The width of the Kondo resonance at the Fermi level
is proportional to the low-energy scale $\omega_{L\sigma} = 
Z_{\sigma}(0)V^{2}$. 
The insets of 
figure~\ref{fig:graph1} show the spectra on a large frequency 
scale. It is seen that for very large frequencies the tails of the spectra 
for all fields are identical which is physically natural since one can 
expect that the effect of the field should dominate only for 
$|\tilde\omega_{\sigma}| \lesssim h\,t_*/(Z_{\sigma}(h)V^{2})$.
Now with increasing magnetic field, the
lattice Kondo resonance splits into two peaks, with the distance between 
the peaks also increasing.
Qualitatively we can understand the results as follows: In the non-interacting
limit for the symmetric case, we have shown that the $c$ and $f$ -levels
shift rigidly due to the Zeeman effect. The same concept is valid
for the asymmetric case also for the non-interacting limit {\it i.e.} 
the spectral function for the up and down spin bands shift rigidly away 
from the Fermi level. 
In the presence of interactions, 
in parallel to the symmetric case, the shift of the spin bands should not be 
rigid due to the competition between Zeeman splitting and Kondo screening. 
And indeed, we find that although the distance between the two peaks varies 
linearly with field in strong coupling (see figure~\ref{fig:shiftdn}), the
slope is not equal to 2 as would have been if the
shift had been rigid (or purely due to Zeeman effect).
In fact, the slope should be equal to 4 as the following argument shows.
From equations ~\eref{eq:rnonc},~\eref{eq:rnonf},~\eref{eq:ref} and 
~\eref{eq:heff}, it is straightforward to see that the splitting
of the Kondo resonance, $S_{KR}$ is given by 
\beq
S_{KR}=\sum_\sigma Z_\sigma(h)\left[h-\sigma\left(\Sigma^R_\sigma(0,h)
-\Sigma^R_\sigma(0,0)\right)\right]
\label{skr}
\eeq
The strong-coupling asymptotic  behaviour of the above quantity
may be inferred using the expressions derived in ~\cite{logan4a}
for the flat-band SIAM. 
\beq
\Sigma^R_\sigma(0,h) - \Sigma^R_\sigma(0,0) = -\frac{4\Delta_0\sigma}{\pi}
\ln\left[\frac{Z_\sigma(h)}{Z_\sigma(0)}\right]
\label{sc1}
\eeq
\beq
\frac{Z_\sigma(h)}{Z_\sigma(0)} = 1+\frac{\pi}{2}\frac{h}{\Delta_0Z_\sigma(0)}
\label{sc2}
\eeq
where $\Delta_0=\pi V^2\rho_0(0)$ is the flat-band hybridization.
Although these were derived for the SIAM, we are justified in
using the same expressions for the PAM for the following reason:
Since the transverse spin-polarization propagator is constructed
using the UHF propagators which do not contain the low energy scale,
and hence are flat for $\tilde{\om}_{h\sigma}\sim{\cal{O}}(1)$,
the strong coupling (SC) asymptotics derived for the SIAM would be similar
to that for the PAM.  The only difference would be in the value of 
$\Delta_0\sim{\cal{O}}(V^2/t_*)$ since the UHF propagators for the PAM 
have a different
structure than that of the SIAM. 
Using equations ~\eref{sc1} and ~\eref{sc2} in ~\eref{skr}, and expanding the
logarithm appearing in equation ~\eref{sc2} to linear order in $h$, we get
\beq
S_{KR}\stackrel{{\rm SC}}{\rightarrow} 4h
\label{eq:skrf}
\eeq
As argued in~\cite{logan4a}, the above result is synonymous
to the Wilson's ratio being equal to 2 in SC.
The inset of figure~\ref{fig:shiftdn} shows the ratio $S_{KR}/h$
as a function of interaction strength. It is seen that even for
$U\sim 2$, the ratio is $\sim 3$, which implies a non-rigid shift
of the spin-bands. Further, with increasing
interaction strength, the ratio increases monotonically. Although
it must asymptotically approach 4, to be consistent with the 
result (equation~\eref{eq:skrf}) obtained above, we are unable to 
access the large $U$ region (for $\ep_c=0.5$) due to the prohibitive
computational expense in handling exponentially small low energy scales.

In a recent DMFT study using QMC
as the impurity solver,  the Kondo lattice model was studied~\cite{beach2}. 
Real frequency
spectra were obtained with a stochastic analytic continuation method.
The parameters chosen were $J=1.6t_*$ and $n_c=0.85$. The main result
was that the bands were found to shift rigidly, in contrast to what we
find above. The difference with our findings could be due to various reasons,
of which the most important seems to be that their low energy scale is 
$0.09t_*$, while our scales are $\sim 10^{-3}t_*$ (see figure 3). This
is significant because only when the Kondo scale is exponentially small
is the coupling between the impurity and the conduction spin renormalized
strongly, and thus the Kondo screening would be strong. If this Kondo
screening is weak, as could probably be in the study mentioned above,
then the Zeeman effect would win easily, and the bands would shift
rigidly.

Now, we turn our attention to field-dependent transport properties. 
At finite temperature, as the spin summed conduction 
electron spectra exhibits universal 
scaling in terms of $T/\om_{L\sigma}$ ($\om_{L\sigma}=Z_{\sigma}(h)V^{2}/t_*$) 
in the strong coupling regime
for low fields, we expect that the resistivity will also exhibit 
scaling in terms of $T/\om_{L\sigma}$. The classic HF metallic resistivity 
increases with temperatures (initially as $T^{2}$), 
 goes through a maximum $T_{max}\simeq \om_L$ (the peak position of $\rho(T)$) and 
then decreases with temperature characteristic of
single impurity behaviour. We observe the same form for the 
resistivity in our calculations.  
\begin{figure}[h]
\centerline{\hbox{
\epsfig{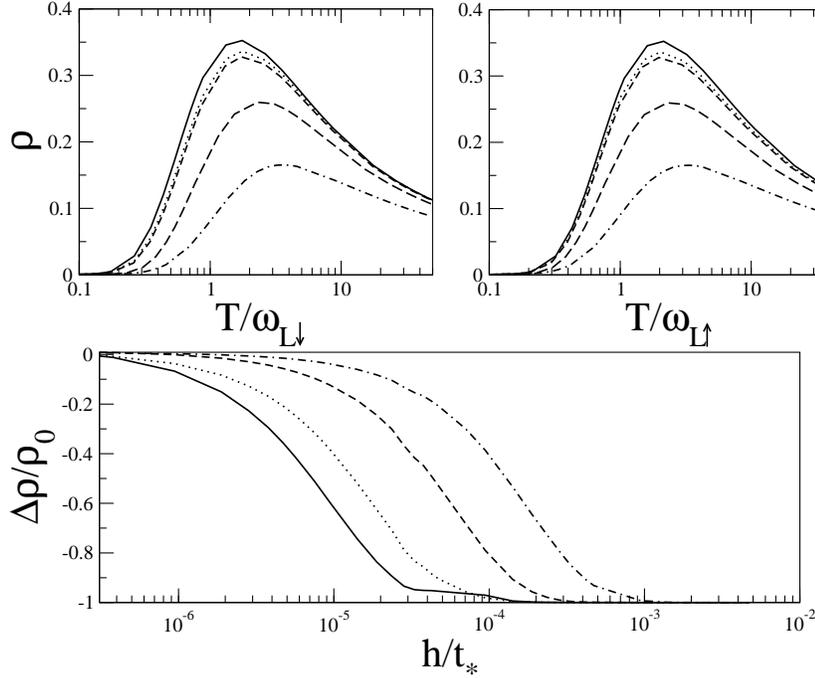}
}}
\caption{Top: Left panel: Resistivity {\it vs} 
$\tilde{\omega}_{h\downarrow}$ for the same parameters, 
but for the various fields ($\tilde{h}=0.0$ (solid), 
0.06 (dotted),
0.1 (dashed), 0.5 (long dashed) and 
2.0 (dot-dashed). 
Right panel: Resistivity {\it vs} $\tilde{\omega}_{h\uparrow}$ for the same
 sets of parameters. Bottom: Magnetoresistivity {\it vs} 
 bare fields $h/t_{*}$ for $U=6.6t_*, \epsilon_{c}=0.3t_*$ and
 $\eta\rightarrow 0$ for $\tilde{T}=0.3$ (solid), 0.6 (dotted),
 2.0 (dashed), 6.0 (dot-dashed).}
\label{fig:graph3}
\end{figure}
In figure~\ref{fig:graph3} (top), the resistivity {\it vs} 
 scaled temperature $T/\omega_{L\sigma}$ is shown for both 
 $\sigma = \uparrow, \downarrow$ and for a range of fields: $\tilde{h}
=0.0$ (solid), 0.06 (dotted), 0.5 (dashed), 
0.5 (long-dashed) and 2.0 (dot-dashed) with $U/t_*$ 
equal to 6.6.
The two most significant features of the above result are --
(i) Negative magnetoresistance is observed over the entire range of 
temperatures as shown in figure~\ref{fig:graph3} (bottom). 
(ii) The maximum in resistivity at the coherence peak reduces in magnitude
and moves to higher temperatures with increasing field.
In previous work, magnetoresistance was found to be positive
for the SIAM~\cite{hewson06} through Fermi liquid theory arguments and for
the Anderson lattice~\cite{tma} through the average T-matrix 
approximation (ATA).
The fundamental difference between the SIAM and the PAM is the presence
of lattice coherence in the latter which manifests itself at low temperatures.
At $T\gtrsim \om_L$, it is clear that magnetic field suppresses spin-flip
scattering (SFS) since the main effect of a field would be to polarize the system.
However, at low temperatures, the physics is different. Kondo singlet 
formation in the SIAM, which at zero field,  quenches
the SFS and hence leads to saturation of resistivity,
is inhibited at finite fields, implying an increase in incoherent scattering
and thus a positive magnetoresistance. In the lattice
case, however, although magnetic field does inhibit the singlet-forming
screening process of local moments, the periodicity
of the local moments (even though not fully screened), introduces lattice 
coherence which strongly
suppresses incoherent scattering at low temperatures, and hence again leads to
negative magnetoresistance. In other words, the incoherent scattering
introduced by magnetic field in the SIAM through the inhibition of
the screening process, is countered in the PAM because of the presence
of lattice/Bloch coherence. It is possible that such an effect is enhanced
in the present calculation due to the use of the DMFT framework, since
non-local dynamical fluctuations are completely neglected
in this framework. The ATA for the Anderson lattice is in a subtle way
somewhat similar to the the single-site DMFT {\em without the self-consistency}.
Which would imply that the effects of lattice coherence are probably
suppressed in the ATA, and hence positive magnetoresistance is observed.
Thus, we conclude that, at higher temperatures ($T\gtrsim \om_L$),
suppression of the spin-flip scattering, and at lower temperatures,
the presence of lattice coherence in the PAM lead to negative 
magnetoresistance in the presence of a field for all $T$.

The shift of coherence peak to higher $T$ with increasing field is
a reflection of the increase in the low energy scale with field (see
figure 3). The
latter result has been obtained  for the SIAM as well ~\cite{logan4a}. 
Since increasing interactions lead to an exponential decrease in $\om_L$,
it appears 
that the polarizing effect of magnetic field is to counter the effect of
interactions, and eventually at very large fields, wipe out local moment
physics (scattering/screening) completely.
At temperatures much higher than the field($T\gg h$), the effect of 
magnetic field is negligible 
implying that the magnetoresistance is almost zero. Next, we compare our theoretical results with 
experiment.

\section{Comparison to $CeB_{6}$}

In this section we want to compare our theoretical 
results with experiment on $CeB_{6}$. The rare-earth hexaboride 
$CeB_6$ has been investigated for many years~\cite{takase,sato,marc,
kimura,nakamura,goodrich}. The cubic-lattice system, at low temperatures,
exhibits various magnetic phase transitions
between 1.6-3.3K, which manifest clearly as kinks in the resistivity,
above which the system is in a paramagnetic phase. It is the $T>$3.3K
phase that we concentrate on, since our approach does not describe
the symmetry broken states. Although there have been extensive studies of this
material, the most detailed magnetoresistance study was carried out by
Takase {\it et al.}~\cite{takase}, who measured the resistivity
of $CeB_{6}$ single crystal in the
temperature range from 3 to 300K and up to magnetic field 85 KOe.
We use their data for comparison to our theory.

\begin{figure}[t]
\centerline{\hbox{
\epsfig{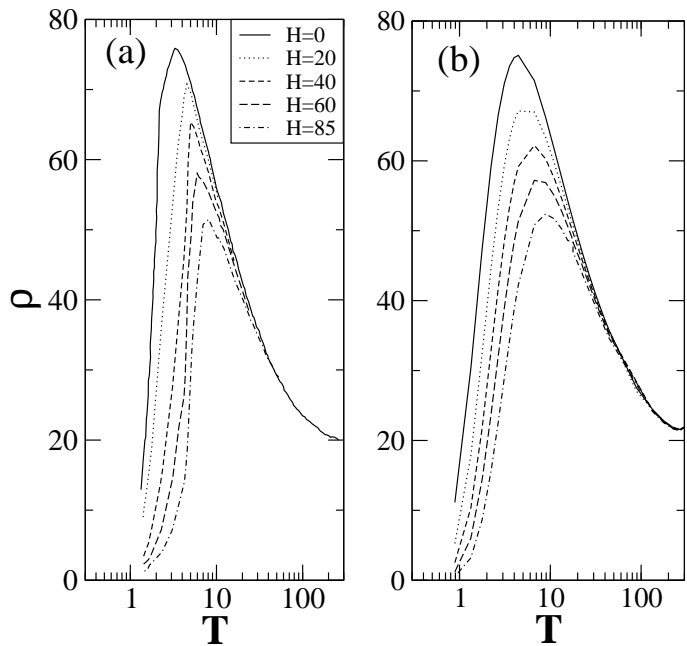}}}
\caption{Resistivity vs temperature of $CeB_{6}$ (a) experiment
(b) theory for the various fields: $H=0$ kOe (solid), 20 kOe (dotted), 
40 kOe (dashed), 60 kOe (long dashed) and 85 kOe (dot dashed).}
\label{fig:graph4}
\end{figure}

From an earlier study~\cite{rajahvexpt}, it is known that the HF system
$CeB_{6}$,
belongs to the moderately strong coupling regime. 
Hence we have taken the theoretical 
results to compare with experiments on $CeB_{6}$ for the 
following parameters: $U=2.4t_*$, $\epsilon_c=0.5t_*$, $\eta \rightarrow 0$ 
and $V^{2}=0.2t_*^2$. 
The experimental resistivity for zero field
shown in the left panel of figure~\ref{fig:graph4}(data from 
Takase {\it et al.}~\cite{takase}) is characteristic of 
classic HF metals. The
resistivity rises sharply from a low value, going through a coherence
peak at $T\sim$4K, and subsequently decreasing through a small log-linear
regime. At higher temperatures~\cite{rajahvexpt,sato}(not shown), resistivity
exhibits single-impurity incoherent behaviour ($T\sim 100-300$K), goes through
a weak minimum ($T\sim 375$K) and finally starts increasing again like a 
normal metal. With increasing magnetic field, negative magnetoresistance
is seen at low temperatures ($T\lesssim 50K$) for all fields. At higher
temperatures, the resistivity remains unaffected by magnetic field.
This behaviour is natural and expected as discussed in section 3. 
The application of magnetic field results in suppression of spin-spin
scattering at $T\gtrsim \om_L$, thus causing a reduction in resistivity. 
At low temperatures lattice coherence takes over and the
magnetoresistance stays negative.
In the strong coupling
regime, the effect of a magnetic
field ($H$) should be expected to extend to temperatures, which are of
a similar magnitude i.e $T\sim g\mu_B H/k_B$, which in the experiment
corresponds to $\sim 5.6K$ for $H=85$kOe (see next paragraph). However,
the highest
field affects the resistivity upto a temperature of $\sim 50K$. Such
behaviour is characteristic of intermediate coupling 
regime~\cite{rajahvtheo}, which is consistent with the model parameters
for CeB$_6$.

To compare the experimental results with our theory, 
two fundamental requirements need to be met. First requirement is that, 
we should extract the contribution to the measured resistivity 
 from phonons($\rho_{ph}(T)$) and
the residual resistivity($\rho(0)$). This is given by, $\rho_{mag}^{exp}(T)=
a(\rho(T)-\rho(0))-\rho_{ph}(T)$ where, a is a constant
which comes from the error of the sample geometry to the
measured resistivity. A detailed discussion of this point is 
given in ref~\cite{rajahvexpt}. In this work, we simply assume $a=1$.
The second is, the value of low-energy scale is needed 
for comparing our theoretical results with experiment. 
For this, we superposed the
theoretical resistivity $\rho_{mag}(T)$ onto $\rho_{mag}^{exp}(T)$ for 
$h=0$ to calculate the value of the low-energy scale. 
The value of the low energy scale turns out to be $\om_L=2.2K$. 
Assuming the $g$-factor to be roughly unity, the magnetic field
of 1kOe can be translated into multiples of the low energy scale.
For $H=1$kOe, $\case{1}{2}g\mu_BH\simeq 0.033K \simeq 0.015 \om_L$. Thus the 
magnetic fields employed in the experiments~\cite{takase} turn out
to be (in multiples of $\om_L$), $\tilde{h}=0, 0.3, 0.6, 0.9$ and $1.25$
corresponding to fields of $H= $ 0, 20kOe, 40kOe, 60kOe and 85kOe respectively.
So, given the model parameters and the low energy scale, along
with the temperature range and the field range, we can compute
the resistivity as a function of temperature at the same fields
as in the experiment.

With a simple multiplicative scaling of the theoretical resistivities on 
the x-axis by the low energy
scale and the y-axis by a single multiplicative factor (same for all fields),
we show the theoretical computed dc resistivities for the same magnetic fields as the
experiment in the right panel of figure~\ref{fig:graph4}.
The experimental resistivity
is found to rise more steeply than the theoretical one. The magnitude
of magnetoresistivity is found to be higher in theory than the experiment.
Nevertheless the functional form agrees quite well. At $T\gtrsim 50K$, the
magnetic field is seen to have very little effect on the resistivity,
both in theory and experiment. Below, 50K, the magnetoresistance in the
theory is negative, which is also in agreement with experiment. The coherence
peak is seen to move to higher $T$ with an increase in field, which
behaviour is seen in the experiment as well. The small log-linear regime
that appears at temperatures higher than the coherence peak temperature
is also reproduced in the theory.
Thus, when compared with the experimental results in figure~\ref{fig:graph4}, 
good qualitative agreement is found between theory and experiment. 

\section{Conclusions.}

In summary, we have employed a non-perturbative local moment approach
to the asymmetric periodic Anderson model
within DMFT in presence of magnetic field. Field-dependent dynamics and 
transport properties of the model
have been computed. In the strong coupling 
Kondo lattice regime of the model, the local $c$- and $f$-electron spectral
functions are found to exhibit universal scaling, being functions solely 
of $\omega/\omega_{L\sigma},T/\omega_{L\sigma}$ ($\omega_{L\sigma}$ being the 
 low-energy scale) for a  given effective field 
$h^{\sigma}_{eff}$. Although the externally applied field is
globally uniform, the effective local field experienced 
by the $c$- and $f$-electrons  differs
because of correlation effects. Fermi liquid behaviour has 
been established even in presence of magnetic fields through 
adiabatic continuity to the non-interacting limit. 
In presence of magnetic field, the quasiparticle peak at $h=0$ and 
$T=0$ splits into two.  The shift of these peaks away from the Fermi 
level is not rigid due to the 
competition between local moment screening 
and Zeeman spin-polarization. Although these 
shifts vary linearly with the field in strong coupling, the slope is enhanced
as compared to the
the non-interacting limit. 
Finally, a comparison of  
theoretical magnetoresistance results with those of CeB$_6$, measured
experimentally, yields good agreement.

\ack

We would like to thank Prof.\ David E.\ Logan for extremely
fruitful discussions, and Department of Science and technology, India
for funding.


\section*{References}
\begin{thebibliography}{99}
\bibitem{grewe} Grewe N and Steglich F 1991 ({\it Handbook on the Physics 
and Chemistry of Rare Earths} vol 14) ed Gschneider K A, Jr.\ and
Eyring L (Amsterdam, Elsevier);
Hewson A C 1993 {\it The Kondo problem to Heavy Fermions},
(Cambridge University Press);
Stewart G R 2001 {\it Rev.\ Mod.\ Phys.} {\bf 73} 797.

\bibitem{vollhard} Vollhardt D 1993 ({\it Correlated Electron Systems} vol 9) 
ed Emery V J (Singapore, World Scientific);
Pruschke T, Jarrell M and Freericks J K 1995 {\it  Adv.\ Phys.} {\bf 44} 187;
Georges A, Kotliar G, Krauth W and Rozenberg M J 1996 {\it
Rev.\ Mod.\ Phys.} {\bf 68} 13;
Gebhard F 1997 {\it The Mott Metal-Insulator Transition}
(Springer Tracts in Modern Physics vol 137) (Berlin: Springer).

\bibitem{qmc} Jarrell M 1995 {\it Phys.\ Rev.\ B} {\bf 51} 7429;
Tahvildar-Zadeh A N, Jarrell M and Freericks J K 1998
{\it Phys.\ Rev.\ Lett.} {\bf 80} 5168; 
Tahvildar-Zadeh A N, Jarrell M, Pruschke T and Freericks J K 1999
{\it  Phys.\ Rev.\ B} {\bf 60} 10782.

\bibitem{nrg}Pruschke T, Bulla R and Jarrell M 2000 {\it Phys.\ Rev.\ B}
{\bf 61} 12799.

\bibitem{ipt} Rozenberg M J, Kotliar G and Kajueter H 1996 
{\it Phys.\ Rev.\ B} {\bf 54} 8452; Vidhyadhiraja N S, 
Tahvildar-Zadeh A N, Jarrell M and Krishamurthy H R 2000
{\it Europhys.\ Lett.} {\bf 49} 459.

\bibitem{nca} Grewe N, Pruschke T and Keiter H 1998
{\it Z.\ Phys.\ B} {\bf 71} 75; Pruschke T and Grewe N 1989
{\it Z.\ Phys.\ B} {\bf 74} 439.

\bibitem{lma_hubb} Logan D E, Eastwood M P and Tusch M A 1997
{\it J.\ Phys.\ Condens.\ Matter} {\bf 9} 4211.

\bibitem{raja_epjb} Vidhyadhiraja N S and Logan D E 2004
{\it Eur.\ Phys.\ J.\ B} {\bf 39}, 313-334.

\bibitem{slave} Newns D M and Read N 1987 {\it  Adv.\ Phys.} {\bf 36} 799;
Sun S J, Yang M F and Hong T M  1993 {\it Phys.\ Rev.\ B} {\bf 48} 16123; 
Burdin S, Georges A and Grempel D R 2000 {\it Phys.\ Rev.\ Lett.} {\bf 85} 1048.

\bibitem{ed} Rozenberg M J 1995 {\it Phys.\ Rev.\ B} {\bf 52} 7369.

\bibitem{pert} Schweitzer H and Czycholl G 1989 Solid State Commun.\ {\bf 69}
179; Schweitzer H and Czycholl G 1991 {\it Phys.\ Rev.\ Lett.} {\bf 67}
3724.

\bibitem{gva} Rice T M and Ueda K 1986 {\it Phys.\ Rev.\ B} {\bf 34}, 6420;
Fazekas P and Brandow B H 1987 {\it Phys.\ Scr.} {\bf 36} 809;
Fazekas P 1987 {\it J.\ Magn.\ Magn.\ Mater.} {\bf 63/64} 545.

\bibitem {tma} Cox D L and Grewe N 1988 {\it Z.\ Phys.\ B} {\bf 71} 321.


\bibitem{logan} Logan D E, Eastwood M P and Tusch M A 1988 
{\it J.\ Phys. Condens.\ Matter} {\bf 10} 2673.

\bibitem{logan1} Glossop M T and Logan D E 2002 {\it J.\ Phys. Condens.\
Matter} {\bf 14} 673;
Dickens N L and Logan D E 2001 {\it J.\ Phys. Condens.\ 
Matter} {\bf 13} 4505.

\bibitem{logan3} Logan D E and Dickens N L 2002 {\it J.\ Phys. Condens.\
Matter} {\bf 14} 3605.

\bibitem{logan4a} Logan D E and Dickens N L 2001 {\it Europhys.\ Lett.} {\bf 54}
227.
\bibitem{logan4b} Logan D E and Dickens N L 2001 {\it J.\ Phys. Condens.\ Matter}
{\bf 13} 9713;
Logan D E and Glossop M T 2002 {\it J.\ Phys. Condens.\ Matter}
{\bf 12} 985;
Glossop M T and Logan D E 2003 {\it J.\ Phys. Condens.\ Matter} {\bf 15} 7519;
Glossop M T and Logan D E 2003 {\it Europhys.\ Lett.} {\bf 61} 810;
Bulla R, Glossop M T, Logan D E and Pruschke T 2000
{\it J.\ Phys. Condens.\ Matter} {\bf 12} 4899;
Smith V E, Logan D E and Krishnamurthy H R 2003 {\it Eur.\ Phys.\ J.\ B} 
{\bf 32} 49. 

\bibitem{logan9} Vidhyadhiraja N S, Smith V E, Logan D E and
Krishnamurthy H R 2003 {\it  J.\ Phys. Condens.\ Matter} {\bf 15} 4045.

\bibitem{rajahvtheo} Logan D E and Vidhyadhiraja N S 2005
 {\it J.\ Phys. Condens.\ Matter} {\bf 17} 2935.

\bibitem{rajahvexpt} Vidhyadhiraja N S and Logan D E 2005
{\it J.\ Phys. Condens.\ Matter} {\bf 17} 2959.

\bibitem {vlma} Kauch A and Byczuk K 2006 {\it Physica B} {\bf 378-380} 297.

\bibitem{saso} Saso T 1997 {\it J.\ Phys.\ Soc.\ Jpn.} {\bf 66} 1175.

\bibitem{beach0} Beach K S D, Lee P A and Monthoux P 2004 {\it Phys.\
Rev.\ Lett.} {\bf 92} 026401.

\bibitem{dp1} Parihari D, Vidhyadhiraja N S and Logan D E 2008 
{\it Phys.\ Rev.\ B} {\bf 78} 035128.


\bibitem{beach1} Viola S K, Beach K S D, Castro Neto A H,
and Campbell D K 2008 {\it Phys.\ Rev.\ B} {\bf 77} 094419. 

\bibitem{beach2} Beach K S D and Assaad F F 2008  {\it Phys.\ Rev.\ B} {\bf 77}
205123.

\bibitem{graf} Graf T, Movshovich R, Thompson J D, Fisk Z and
Canfield P C 1995 {\it Phys.\ Rev.\ B} {\bf 52}, 3099.

\bibitem{chen} Chen C, Li Z -Z and Xu W  1993 {\it J.\ Phys.\ Condens.\ Matter} 
{\bf 5} 95; Chen C and Li Z -Z 1994 {\it J.\ Phys.\ Condens.\ Matter} {\bf 6}
2957.

\bibitem{luttinger} Luttinger J M and Ward J C 1960 {\it Phys.\ Rev.}
{\bf 118} 1417.

\bibitem{hewson06} Hewson A C, Bauer J and Koller W 2006 {\it Phys.\ Rev.\ B}
{\bf 73} 045117.

\bibitem{takase} Takase A, Kojima K, Komatsubarai T and Kasuya T 1980
{\it Sol.\ St.\ Comm.} {\bf 36}  461.

\bibitem{sato} Sato N, Sumiyama A, Kunii S, Nagano H and Kasuya T 1985
{\it J.\ Phys.\ Soc.\ Japan} {\bf 54} 1923.

\bibitem{marc} Marcenat C, Jaccard D, Sierro J, Floquet J,
Onuki Y and Komatsubahara T 1990  {\it J.\ Low Temp.\ Phys.} {\bf 78} 261.

\bibitem{kimura} Kimura S I, Nanba T, Kunii S and Kasuya T 1994
{\it Phys.\ Rev.\ B} {\bf 50} 1406 (1994).

\bibitem{nakamura} Nakamura S, Goto T and Kunii S 1995 {\it J.\ Phys.\ Soc.\
Japan} {\bf 64} 3491.

\bibitem{goodrich} Goodrich R G {\it et al.} 2004 {\it Phys.\ Rev.\ B} {\bf 69}
054415.


\end{thebibliography}
\end{document}